\documentclass[lettersize,journal]{IEEEtran}
\usepackage{amsmath,amsfonts}
\usepackage{algorithmic}
\usepackage{algorithm}
\usepackage{array}
\usepackage{textcomp}
\usepackage{stfloats}
\usepackage{url}
\usepackage{verbatim}
\usepackage{graphicx}
\usepackage{cite}
\usepackage{bm}
\usepackage{xspace}
\usepackage[colorlinks=true, linkcolor=blue, citecolor=blue, urlcolor=blue]{hyperref}
\usepackage{cite}
\usepackage{amsmath,amssymb,amsfonts}
\usepackage{algorithmic}
\usepackage{graphicx}
\usepackage{textcomp}
\usepackage{xcolor}
\def\BibTeX{{\rm B\kern-.05em{\sc i\kern-.025em b}\kern-.08em
    T\kern-.1667em\lower.7ex\hbox{E}\kern-.125emX}}

\usepackage{tikz}
\usepackage{subcaption} 
\usepackage{mdframed} 
\usepackage[most]{tcolorbox}

\author{
Shvetank Prakash\textsuperscript{$\ast$}
Andrew Cheng\textsuperscript{$\ast$}
Jason Yik\textsuperscript{$\ast$}
Arya Tschand\textsuperscript{$\ast$}
Radhika Ghosal\textsuperscript{$\ast$}\\[.2em]
Ikechukwu Uchendu\textsuperscript{$\ast$}
Jessica Quaye\textsuperscript{$\ast$}
Jeffrey Ma\textsuperscript{$\ast$}
Shreyas Grampurohit\textsuperscript{$\S$}
Sofia Giannuzzi\textsuperscript{$\ast$}\\[.2em]
Arnav Balyan
Fin Amin\textsuperscript{$\gamma$}
Aadya Pipersenia\textsuperscript{$\S$}
Yash Choudhary\textsuperscript{$\S$}
Ankita Nayak\textsuperscript{$\phi$}\\[.2em]
Amir Yazdanbakhsh\textsuperscript{$\dagger$}
Vijay Janapa Reddi\textsuperscript{$\ast$}\\[.4em]
\it 
\textsuperscript{$\ast$}Harvard University %
\textsuperscript{$\S$}Indian Institute of Technology Bombay\\[0.2em] %
\textsuperscript{$\gamma$}North Carolina State University %
\textsuperscript{$\phi$}Qualcomm AI Research
\textsuperscript{$\dagger$}Google DeepMind
\vspace{-1.0em}
}

\newcommand{\ours}{{QuArch}\xspace}
\newcommand{\oursnospace}{{QuArch}}

\begin{document}

\newmdenv[
  backgroundcolor=gray!10,
  linecolor=black,
  linewidth=1pt,
  roundcorner=5pt,
  innertopmargin=10pt,
  innerbottommargin=10pt
]{example}

\newcommand{\ballnumber}[1]{\tikz[baseline=(myanchor.base)] \node[circle,fill=.,inner sep=1pt] (myanchor) {\color{-.}\bfseries\footnotesize #1};}

\title{\ours: A Question-Answering Dataset for \\AI Agents in Computer Architecture}
\maketitle
\vspace{-10.0em} 


\begin{abstract}
We introduce \ours, a dataset of 1500 human-validated question-answer pairs designed to evaluate and enhance language models' understanding of computer architecture. The dataset covers areas including processor design, memory systems, and performance optimization. Our analysis highlights a significant performance gap: the best closed-source model achieves 84\% accuracy, while the top small open-source model reaches 72\%. We observe notable struggles in memory systems, interconnection networks, and benchmarking. Fine-tuning with \ours improves small model accuracy by up to 8\%, establishing a foundation for advancing AI-driven computer architecture research. The dataset and leaderboard are at \url{https://harvard-edge.github.io/QuArch/}.
\end{abstract}
\section{Introduction}
Generative Artificial Intelligence (GenAI) has transformed domain-specific tools across diverse fields such as medicine, mathematics, law, finance, and software engineering~\cite{chen2024overview}.
In contrast, hardware engineering has lagged significantly in adopting AI-driven solutions. 
This gap is evident in both the limitations of current language models (LMs) and the scarcity of specialized datasets tailored for hardware.
For instance, engineering tasks often perform poorly on general benchmarks like MMLU-Pro~\cite{wang2024mmlu}, highlighting the inadequacy of existing models in understanding domain-specific intricacies. 
While electronic design automation (EDA) has seen recent progress with datasets for tasks such as register-transfer level (RTL) generation~\cite{verilogeval,rtl3}, security analysis~\cite{securitybug1}, and verification~\cite{nl2spec}, computer architecture remains underrepresented. 
Without resources to benchmark and advance AI models, the field is limited in its ability to improve AI-driven solutions.
Datasets play a key role in enabling AI agents.
While general-purpose datasets provide broad knowledge, domain-specific datasets are indispensable for developing expertise in areas like computer architecture. 
These targeted datasets enable AI models to not only demonstrate foundational understanding but also tackle advanced problem-solving tasks within specific domains~\cite{rogers2023qa}. 
Mastery of domain knowledge is a prerequisite for sophisticated reasoning~\cite{duncan2007role, krieger2004domain}. 
In architecture, such proficiency is essential for developing practical AI-driven tools and agents~\cite{damaniwarpdrive}.
Without a deep understanding of core concepts—such as processor execution, memory hierarchy, and parallelism—it becomes impossible to conduct analyses of the complex trade-offs inherent in system design.
\begin{figure}[ht]
    \centering
    \begin{tcolorbox}[
        colframe=blue!70!black, 
        colback=blue!5!white,   
        arc=2mm,                  
        boxrule=0.5mm,            
        width=0.488\textwidth,    
        left=1mm,                 
        right=1mm,                
        top=1mm,                  
        bottom=0mm                
    ]
            \small  
            \underline{{\textbf{Example 1}}}\\[0.5em]
            \textbf{Topic:} Processor Architecture\\
            \textbf{Q:} In \texttt{\textunderscore\textunderscore\textunderscore}, each processor has its own local memory system. \\
            \textbf{A:} (a) symmetric multiprocessing (b) asymmetric multiprocessing (c) core-based multiprocessing \textbf{(d) clustered multiprocessing} \\
            
            \underline{{\textbf{Example 2}}}\\[0.5em]
            \textbf{Topic:} Storage Systems\\
            \textbf{Q:}  Moving compute closer to the \texttt{\textunderscore\textunderscore\textunderscore} in solid state drives (SSDs) offers higher bandwidth but introduces challenges in managing frequent errors. \\
            \textbf{A:} (a) controller \textbf{(b) NAND dies} (c) cache (d) DRAM \\

            \underline{{\textbf{Example 3}}}\\[0.5em]
            \textbf{Topic:} Architectural Support\\
            \textbf{Q:} \texttt{\textunderscore\textunderscore\textunderscore} translates the logical address into a physical address. \\
            \textbf{A:} \textbf{(a) MMU} (b) Translator (c) Compiler (d) Linker \\ 
    \end{tcolorbox}
    \caption{Example QAs from \ours for various topics curated from different sources. The bolded answer is correct.}
    \label{fig:samples_qas}
    \vspace{-1.5em}
\end{figure}

To address this challenge, we introduce \ours (pronounced 'quark')---the first \textbf{Qu}estion-Answering Dataset specifically tailored for Computer \textbf{Arch}itecture. 
This dataset addresses a critical gap in evaluating LMs understanding of architectural concepts. 
\ours comprises 1,500 rigorously curated question-answer pairs, manually annotated by domain experts. 
It spans both foundational computer architecture principles and contemporary topics, such as deep learning accelerators and quantum computing architectures.

\ours is designed to assess an LMs' ability to retrieve and apply domain-specific knowledge, a prerequisite for addressing advanced problem-solving challenges.
The dataset serves as a benchmark for both theoretical understanding and practical application in computer architecture.

Leveraging \ours, we provide the first comprehensive evaluation of the architectural knowledge encoded in state-of-the-art (\textsc{SoTA}) LMs.
Our results show LM accuracies ranging from 39\% to 84\%, revealing a significant 12\% knowledge gap between the best-performing small open-source \textsc{LM} and large closed-source \textsc{LM}. 
Additionally, our experiments demonstrate the utility of \ours in fine-tuning small \textsc{LM}s, yielding performance improvements of 5.4\%–8.3\%.
\section{Related Work}
Recent efforts have explored the use of GenAI in hardware design. 
For example, NVIDIA's ChipNeMo~\cite{chipnemo} introduced foundation models tailored for chip design tasks, while other studies have focused on developing LM-based tools for RTL generation~\cite{rtl3} and hardware verification~\cite{nl2spec}. 
Additionally, evaluation datasets in this domain have been created for specific implementation tasks, such as VerilogEval~\cite{verilogeval} for RTL generation, and general-purpose benchmarks like MMLU~\cite{mmlu}, which assess engineering knowledge broadly across disciplines.
However, none of these prior works specifically evaluate LM understanding of computer architecture concepts. 
This critical gap limits the ability to assess and advance LM capabilities for architectural challenges. 
\ours addresses this unmet need by introducing a focused question-answering dataset specifically designed to evaluate architectural knowledge (Figure~\ref{fig:samples_qas}). 
The dataset combines synthetic data generation~\cite{shakeri2020} with rigorous expert validation, ensuring both comprehensive coverage and high-quality questions.

\section{\ours}
In this section, we discuss the construction and characteristics of the first version of the dataset: \ours v0.1.
\begin{figure}[t]
    \centering
    \includegraphics[width=\linewidth]{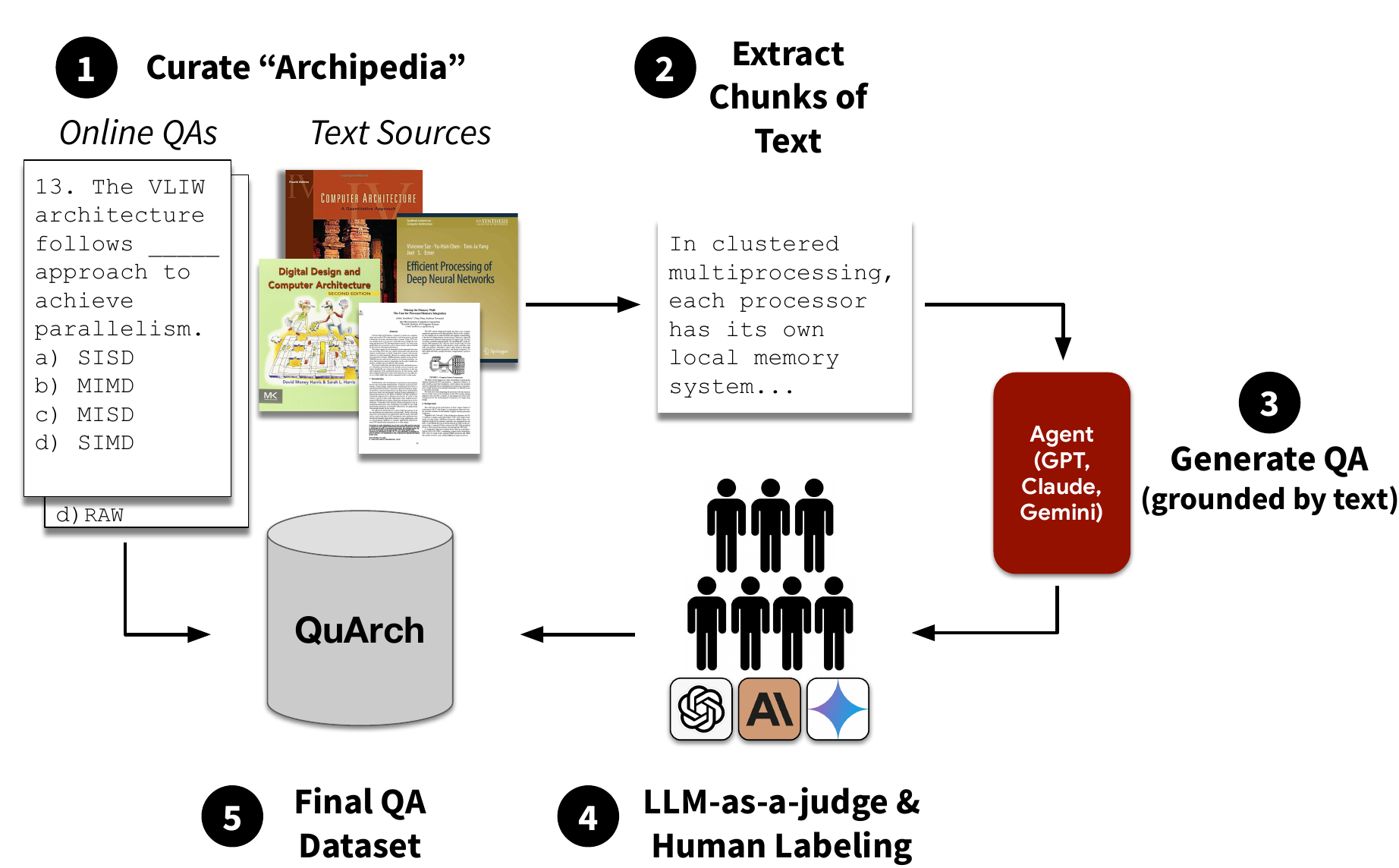}
    \caption{\ours dataset construction pipeline.}
    \label{fig:quarch_sdg}
    \vspace{-1.5em}
\end{figure}

\subsection{Dataset Curation: The Archipedia Corpus}
The construction of \ours follows a systematic process, as depicted in Figure~\ref{fig:quarch_sdg}. 
We first curated ``Archipedia,''\footnote{Data was downloaded and evaluated solely by Harvard University.} a term we use to describe a comprehensive compilation of computer architecture knowledge that was assembled for this work. 
Archipedia synthesizes five decades of information, drawing from academic literature, educational materials, technical documentation, and industry sources across the computing landscape. 
This extensive corpus captures the evolution of the field over the past 50 years, incorporating contributions from leading institutions, researchers, and organizations globally. 
Currently, the corpus exceeds 1 billion tokens in size.
Archipedia covers the full spectrum of computing systems, from foundational topics in computer architecture to cutting-edge technologies. 
It includes domains such as VLSI design and technology, embedded systems and IoT, parallel and distributed processing, hardware-software co-design, and design automation.
In addition, the corpus integrates specialized areas such as computer-aided design tools, hardware security, and quantum computing. 
To ensure comprehensive coverage, this resource is further enriched with advanced lecture materials and thus provides a diverse and balanced knowledge base.
\begin{figure}[t]
    \centering
    \includegraphics[width=\linewidth]{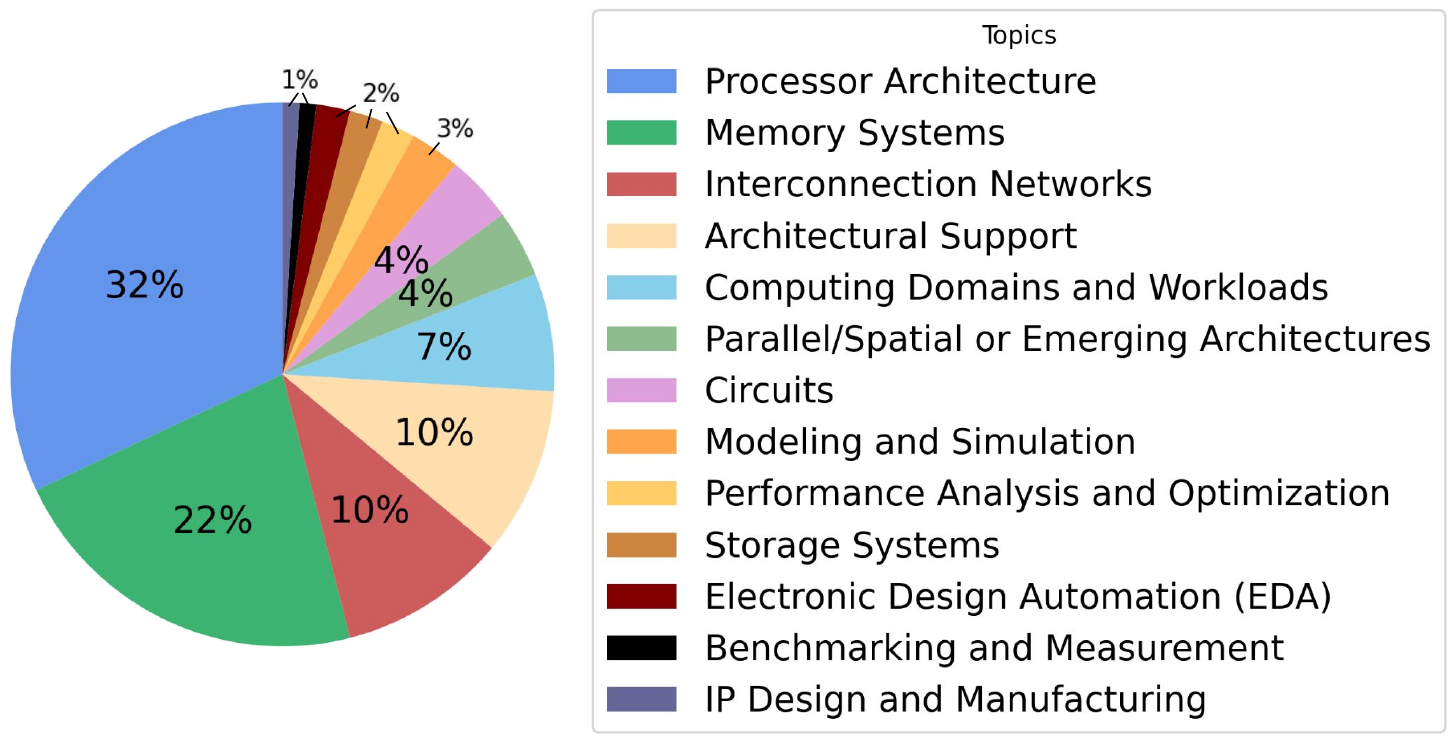}
    \caption{Distribution of computer architecture topics in \ours.}
    \label{fig:quarch_topics}
    \vspace{-1.5em}
\end{figure}

\subsection{Dataset Generation: QA Creation}
\label{sec:quarch_generation}
The knowledge curation phase (Steps \protect\ballnumber{1} and \protect\ballnumber{2} in Figure~\ref{fig:quarch_sdg}) established a foundation of well-accepted architectural concepts and principles by leveraging diverse sources. 
This effort utilized the Archipedia corpus, which provided a comprehensive resource for generating questions for \ours.

In the QA generation phase (Step \protect\ballnumber{3}), commercial LMs were used to synthesize questions grounded in the academic content of Archipedia to ensure technical rigor. 
The LMs were tasked with creating cloze-style multiple-choice QAs~\cite{rogers2023qa} to balance educational value with practical assessment.

The validation phase (Step \protect\ballnumber{4}) involved a multi-tiered review process that combined human expertise with LM assistance. 
Questions derived from undergraduate-level sources were reviewed by an expert with graduate-level architectural expertise, supplemented by LM validation using the ``LLM-as-a-judge'' technique~\cite{zheng2023judging}. 
Advanced topics were evaluated by a pool of eight experts, and QAs were independently validated by three reviewers who reached consensus to ensure accuracy.

To further enhance the validation process, human experts and LM reviewers received contextual fragments of the source text, transforming the task into a focused reading comprehension exercise. 
This approach enabled the identification and removal of questions lacking definitive answers or those too narrowly scoped for meaningful assessment. 
The validation process, facilitated through the Label Studio platform~\cite{labelstudio}, ensured the resulting dataset effectively tests both foundational principles and complex system trade-offs.
The final dataset (Step \protect\ballnumber{5}) supplemented these expert-validated questions with additional QAs freely available from an online education platform, enriching the dataset's depth and coverage.

\subsection{Dataset Coverage: Architecture Topics}
\ours v0.1 contains 1,547 question-answer pairs. It captures the breadth of architecture in 13 core areas derived from key themes of the past decade (Figure~\ref{fig:quarch_topics}).
Processor architecture accounts for the largest proportion of questions (32\%), followed by memory systems (22\%) and interconnection networks (10\%). 
The topic distribution was determined through a two-stage classification process using OpenAI's \texttt{text-embedding-3-large} model~\cite{openaiembeddings}. 
In the first stage, the word embedding model generates vector representations of topics and questions, and cosine similarity~\cite{rahutomo2012semantic} is used to identify the three most relevant topics for each question. 
In the second stage, an LM selects the final topic from these candidates for accurate categorization and scalability.

Figure~\ref{fig:samples_qas} illustrates the diversity and depth of architectural concepts covered in \ours.
Foundational questions assess core principles of processor architecture, such as ``In clustered multiprocessing, each processor has its own local memory system.'' 
More advanced questions probe emerging trade-offs, exemplified by Question \#2 in Figure~\ref{fig:samples_qas}, which addresses near-storage computing: Moving compute closer to NAND dies in solid-state drives increases bandwidth while mitigating the risk of silent data corruption. 
Finally, Question \#3 highlights the dataset’s comprehensive scope by including critical system-level concepts, such as virtual memory and address translation.

\begin{figure}[t]
    \centering
    \includegraphics[width=\linewidth]{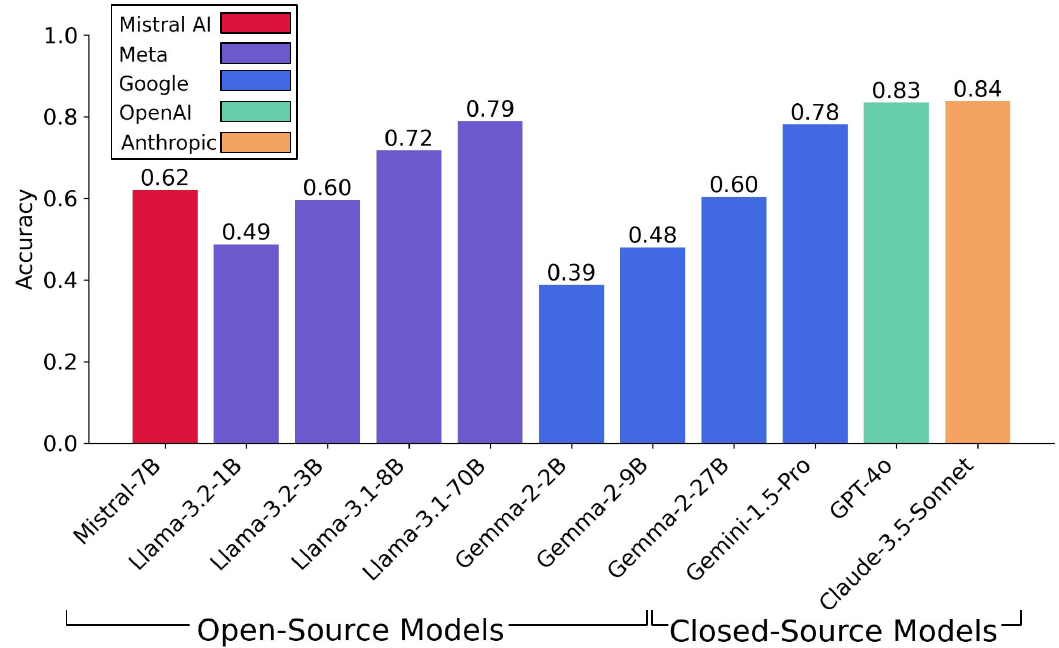}
    \caption{\ours accuracy ranges from 39\%-84\%. Larger models ($>$70B parameters) attain a max of 84\%. Small model ($<$10B parameters) performance  drops 12\% in comparison.}
    \label{fig:baseline_model_accuracy}
    \vspace{-1.5em}
\end{figure}

\section{Results}
To evaluate the state of computer architecture knowledge embedded in LMs, we assess knowledge retrieval capabilities of \textsc{SoTA} models and explore opportunities to improve them.

\subsection{Experimental Setup}
We evaluated both open-source and closed-source language models on \ours. 
The evaluation included large-scale models such as GPT-4o, Claude-3.5 Sonnet, and Gemini-1.5 Pro, as well as open-source models with parameter counts ranging from 1B to 70B from Google, Meta, and Mistral AI. 
Each model was presented with questions in a multiple-choice format, requiring the selection of the correct answer from four options. 
Models were prompted to ``act as computer architecture experts'' and were evaluated in a zero-shot setting, with no additional context provided beyond the questions themselves. 
This setup was designed to test the models' baseline understanding of architectural concepts.
Accuracy (percentage of correct answers) served as the primary evaluation metric.

\begin{figure}[t]
    \centering
    \includegraphics[width=\linewidth]{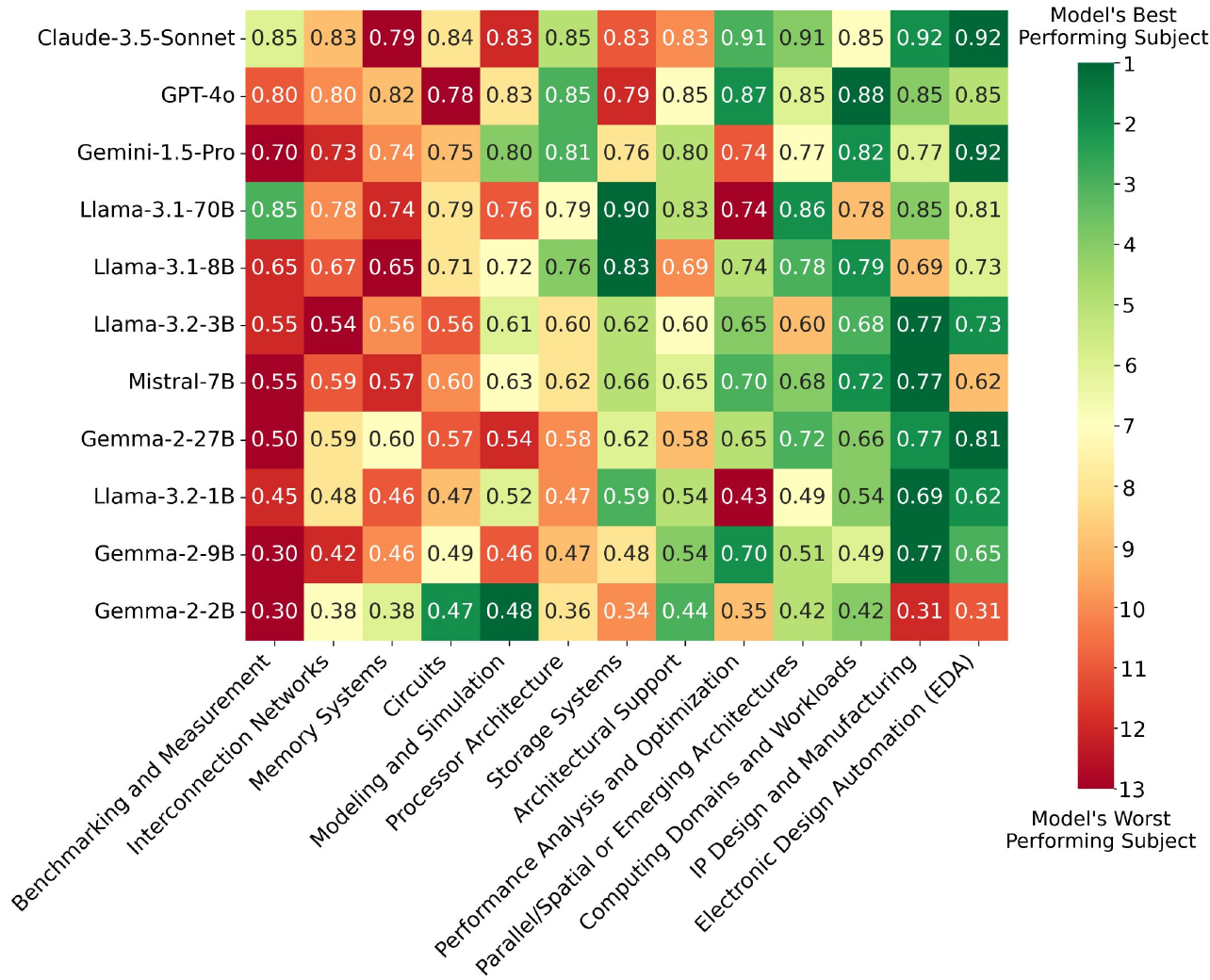}
    \caption{Performance breakdown across topics. Color intensity indicates topic's relative (intra-model) performance, with darker green showing stronger understanding and darker red showing weaker areas. Memory systems and interconnects are more challenging for current LMs.
    Benchmarking also shows low performance but only accounts for 1\% of the QAs.
    }
    \label{fig:topic_accuracy_heat_map}
    \vspace{-1.5em}
\end{figure}

\subsection{Understanding of Architecture Concepts}

Figure~\ref{fig:baseline_model_accuracy} presents the baseline performance of LMs on \ours. 
The top-performing model achieves 84\% accuracy, reflecting a relatively strong but incomplete understanding of architecture concepts. 
In particular, a substantial knowledge gap exists: the best-performing small open-source LM ($<$10B parameters) underperforms by 12\% on the same questions.
The observed performance ceiling of 84\% suggests that current LMs still have significant room for improvement in understanding the fundamentals of computer architecture concepts.
These findings have important implications for the development of agentic tools for hardware design~\cite{damaniwarpdrive}. 
While current models exhibit a reasonable grasp of basic architectural concepts, they may require supplementary support or verification mechanisms when addressing complex system-level decisions.

\subsection{Analysis by Architecture Topics}
Figure~\ref{fig:topic_accuracy_heat_map} presents a heatmap illustrating LM performance across various architecture topics. 
Each cell contains the raw accuracy values for a specific topic and corresponding model. 
To account for substantial differences in raw accuracy due to varying model capacities, the heatmap employs color gradients to represent performance \textit{relative to each model's overall accuracy}.
Dark green denotes the strongest performance for a given model, while dark red highlights the weakest, offering a clearer perspective on relative strengths and weaknesses.

The analysis reveals distinct patterns in how LMs comprehend different architecture topics. 
Models exhibit their strongest performance in topics such as EDA concepts, IP design and manufacturing, parallel processing architecture fundamentals, and compute workload characterization (relative to other topics).
Conversely, significant challenges are observed in three critical areas: memory systems, interconnection networks, and benchmarking and measurement. 
These weaknesses are consistent with expectations, as questions in these areas often involve nuanced system-level interactions and intricate technical trade-offs.
Although general trends are evident, individual model performance varies considerably.
For example, Llama-3.1-70B demonstrates notable strength in storage systems QAs, outperforming larger models such as Claude 3.5 Sonnet and GPT-4o.
Furthermore, it performs well in benchmarking and measurement---a topic that accounts for only 1\% of the dataset (Figure~\ref{fig:quarch_topics})---unlike most other models.
These findings underscore specific gaps in the architectural knowledge of current LMs. 
Such disparities likely stem from differences in the data blends used during model training. 
For instance, the stronger performance in parallel processor architectures, such as GPUs and deep learning accelerators, likely reflects increased community focus and the abundance of academic text on these topics.
In contrast, the weaker performance in memory systems and interconnection networks suggests these complex, system-level concepts warrant greater emphasis when developing future AI-based tools for architects.

\subsection{\ours as an Architecture Benchmark}
\label{sec:quarch_as_benchmark}
To validate the effectiveness of \ours as a benchmark, we evaluate its ability to distinguish between the capabilities of different LMs and compare it to established QA benchmarks. 
An effective benchmark should strike a balance---it should neither be trivial to solve nor overly challenging so that it's unattainable. 
\ours satisfies this criterion, as even the top-performing models, such as GPT-4o and Claude 3.5-Sonnet, achieve accuracies of only 83–84\%. 
This highlights substantial room for improvement in architectural understanding.
\begin{table}[t!]
\centering
\resizebox{\columnwidth}{!}{
\begin{tabular}{|l||c|c||c|}
\hline
\textbf{Model} & \textbf{MMLU (\%)} & \textbf{GPQA (\%)} & \textbf{\ours (\%)} \\
\hline
GPT-4o           & 88.7 & 53.6 & 83.0 \\
Claude 3.5 Sonnet & 88.3 & 59.4 & 84.0 \\
\hline
\end{tabular}
}
\caption{\textsc{SoTA} QA benchmark accuracy versus \ours.}
\label{tab:qa_benchmarks_transposed}
\vspace{-1.5em}
\end{table}

This performance ceiling is consistent with what is observed on other well-established QA benchmarks. 
As shown in Table~\ref{tab:qa_benchmarks_transposed}, the same models achieve comparable performance (88–89\%~\cite{openai2024gpt4o, claude_3_5_sonnet}) on general knowledge benchmarks like MMLU\cite{mmlu}, which include engineering-related QAs. 
This suggests that \oursnospace’s difficulty aligns well with other technical assessments.
In contrast, GPQA~\cite{gpqa}, one of the most challenging benchmarks available, achieves lower accuracies (54–59\%) due to its hand-crafted questions that require advanced QA skills beyond knowledge retrieval. 
\oursnospace’s positioning between MMLU and GPQA demonstrates its value as a meaningful and balanced measure of model capabilities.
Furthermore, the room for improvement, particularly in advanced architecture topics, highlights \oursnospace’s potential to track progress in LMs’ understanding of computer architecture. 
However, further expansion and refinement of the dataset will be necessary to fully realize its benchmarking potential.

\subsection{\ours as an Architecture Training Dataset}
ML datasets serve dual purposes: benchmarking and training. 
In this section, we investigate whether \ours can enhance the domain-specific knowledge of LMs through fine-tuning. 
To this end, we fine-tuned instruction-tuned variants of small open-source LMs using an 80-20 train-test split. 
To ensure robustness in the training evaluation, we employed repeated random train-test splitting with five different seeds. 

Table~\ref{tab:quarch_fine_tuning} reports mean test set accuracy improvements after fine-tuning on \ours across different train-test splits. 
The results indicate significant performance gains, even with the relatively small size of the dataset. 
On average, instruction-tuned variants of Gemma-2-2B and Llama-3.2-3B demonstrated improvements ranging from 5.4\% to 8.3\%. 
These substantial gains underscore the potential of \ours to enhance LMs’ understanding of computer architecture.
Moreover, the results highlight the importance of developing larger, diverse datasets to further advance AI-based solutions in this domain.

\section{Conclusion}
\ours is the first question-answering dataset for computer architecture, providing a means to evaluate domain knowledge. 
Through \ours, we uncover both the strengths and limitations of SoTA LMs to reveal substantial room for improvement in this domain.
\ours benchmarks knowledge retrieval---an essential foundation for advancing the integration of AI in computer architecture---but future datasets must build on \ours to evaluate more complex capabilities, including advanced reasoning, system-level planning, and architectural design. 
Realizing these goals will require large-scale collaboration between academia and industry, ensuring AI tools for architecture evolve to meet the field's growing demands.
\begin{table}[t!]
\centering
\resizebox{\columnwidth}{!}{
\begin{tabular}{|l||c|c||c|}
\hline
\textbf{Model} & \textbf{Original (\%)} & \textbf{Fine-Tuned (\%)} & \textbf{Improvement (\%)} \\
\hline
Gemma-2-2B      & 38.7 $\pm$ 3.0           & 47.0 $\pm$ 3.0          & +8.3               \\
Llama-3.2-3B    & 59.6 $\pm$ 1.0           & 65.0 $\pm$ 2.0          & +5.4               \\
\hline
\end{tabular}
}
\caption{Mean and standard deviation of test accuracy when fine-tuning on \ours using repeated random train-test splits.}
\label{tab:quarch_fine_tuning}
\vspace{-1.5em}
\end{table}
\section{Acknowledgements}
We would like to acknowledge and thank the many students from around the world who were instrumental in the early development efforts of \ours: 
Timothy Akert, Adolfo Balderas, Nandini Bhattad, Rushi Chavda, Arjun Chitla, Anjali Choudhary, Hardik Jagga, Satyapragnya Kar, Subash Katel, Anirvan Krishna, Ujjwal Kumar, Sushant Kumar, Shrestha Mishra, Vishnu Nand, Andrew Ogundimu, Arianna Ording, Elsa Oreen, Aarya Pakhale, Umair Paranjpye, Daksh Parikh, Haresh Perera, LuzZelenia Perez-McNeill, Frankie Francisco Pinaminjarez, Arnav Raj, Debarpita Saha, Anmol Saraf, Fatima Shah, Akshit Sharma, Aishani Singh, Hartej Soin, Yurun Song, Akarsh Srivastava, Samuel Stankiewicz, Keerthana Subramanian, Kavya Subramanian, Sujay Suribhotla, Aman Tyagi, Maneesh Vaddi, Ankit Walishetti, Judong Wang, Max Zhang, and Junchen Zhao.
We also extend our gratitude to Derek Lockhart and James Laudon for their valuable feedback on this paper.
Finally, we would like to thank Christos Kozyrakis for the use of a qualifying exam question that helped motivate the need for a dataset assessing foundational domain knowledge.

\bibliographystyle{IEEEtran}
\bibliography{refs.bib} 

\end{document}